\RequirePackage{fix-cm}
\documentclass[twocolumn]{svjour3}          
\smartqed  
\usepackage{graphicx}
\usepackage{latexsym}
\usepackage{amsmath}
\usepackage{amssymb}
\usepackage{multirow}
\usepackage{cuted}

\journalname{Brazilian Journal of Physics}

\begin{document}

\title{Computing partial transposes and related entanglement functions}

\author{Jonas Maziero}
\institute{Jonas Maziero \at Departamento de F\'isica, Centro de Ci\^encias Naturais e Exatas, Universidade Federal de Santa Maria, Avenida Roraima 1000, 97105-900, Santa Maria, RS, Brazil \\ Instituto de F\'isica, Facultad de Ingenier\'ia, Universidad de la Rep\'ublica, J. Herrera y Reissig 565, 11300, Montevideo, Uruguay \\ \email{jonas.maziero@ufsm.br}}

\date{Received: date / Accepted: date}

\maketitle

\begin{abstract}
The partial transpose (PT) is an important function for entanglement
testing and quantification and also for the study of geometrical aspects
of the quantum state space. In this article, considering general bipartite
and multipartite discrete systems, explicit formulas ready for the
numerical implementation of the PT and of related entanglement functions
are presented and the Fortran code produced for that purpose is described.
What is more, we obtain an analytical expression for the Hilbert-Schmidt
entanglement of two-qudit systems and for the associated closest separable
state. In contrast to previous works on this matter, we only use the
properties of the PT, not applying Lagrange multipliers.
\end{abstract}


\maketitle

\section{Introduction}

\label{intro}

The correlations among the constituent particles of physical systems
are of central importance for science \cite{Mermin}. In quantum information
science (QIS) \cite{Nielsen=000026Chuang,Preskill,Wilde}, there are
several types of correlations \cite{Winter_corr,Horodecki_RMP,Lucas_review,Lucas_PTRSA,Wehner_BellNL,Cavalcanti_Steering}.
Entanglement is one kind of quantum correlation, one which is widely
recognized as being the fuel for the more efficient realization of
several information manipulation tasks \cite{Leuchs_Eapps,Diogo_dipolar,Gisin_QK,Maccone_Met}. 

Entanglement quantifiers (EQs) are functions which are null only for
states that can be prepared using local quantum operations and classical
communication (the separable states) and which do not increase under
such kind of transformation \cite{Horodecki_RMP}. Nowadays, there
are several proposals of EQs in the literature \cite{Plenio_Erev,Davidovich}.
One common feature of these quantities is that they are very hard
to compute analytically in the general case \cite{Gharibian,Huang}.
This motivates the consideration of entanglement functions (EFs),
which possess some, but not all, of the properties one may request
for a good EQ. The partial transposition (PT) provides the most famous
and tractable separability criterion and EFs \cite{Peres_PT,Horodecki_PT,Werner_En,Plenio_Eln},
and is relevant for a myriad of investigations in QIS.

In this article, we present a detailed description of the partial
transposition map and of related EFs. In addition to that, we obtain
an analytical formula for the Hilbert-Schmidt entanglement (HSE) of
two-qudit\footnote{A qudit is a $d$-level quantum system.} systems
and for the associated nearest separable state. We also describe Fortran
code\footnote{The Fortran code used in this article is part of the Fortran Library
for Quantum Information Science and can be accessed freely in: https://github.com/jonasmaziero/LibForQ. For the description of some related tools, see Refs. \cite{Maziero_probs,Maziero_rhos,Maziero_libforro,Maziero_GM,Maziero_PTr}.} produced to compute all the functions regarded here.

The remainder of this article is structured as follows. In the next
section, we start introducing the transposition map (Sec. \ref{transp}).
In the sequence we use it to discuss the partial transposition operation
in the contexts of bipartite (Sec. \ref{sec:PT2}) and of multipartite
discrete systems (Sec. \ref{sec:PT3}). In Sec. \ref{Peres}, we recall
the Peres' separability criterion and some related EFs. The analytical
calculation of the HSE is addressed in Sec. \ref{sec:HSE}. Some final
remarks and open questions are included in Sec. \ref{conc}. 

\section{Partial transposes and related entanglement functions}

\label{sec:PT}

\subsection{The transposition map}

\label{transp}

Before introducing the partial transposition map, let's discuss the
transposition operation. Let $\rho$ be a general linear operator
defined on the Hilbert space $\mathcal{H}$. Let $\{|j\rangle\}_{j=1}^{d}$
be an orthonormal basis for $\mathcal{H}$, with $d=\dim\mathcal{H}$.
Then we can write the matrix representation: $\rho={\textstyle \sum_{j,k=1}^{d}}\langle j|\rho|k\rangle|j\rangle\langle k|.$
In the sequence, $|j\rangle$ is assumed to be the standard computational
basis in $\mathcal{H}$. By definition, the \emph{transposition} map
$T$ is linear and acts on the computational basis as follows:
\begin{equation}
T(\sum_{j,k}c_{jk}|j\rangle\langle k|):=\sum_{j,k}c_{jk}T(|j\rangle\langle k|):=\sum_{j,k}c_{jk}|k\rangle\langle j|,\label{eq:Tmap}
\end{equation}
with $c_{jk}\in\mathbb{C}$. Thus,
\begin{eqnarray}
T(\rho) & = & T(\sum_{j,k=1}^{d}\langle j|\rho|k\rangle|j\rangle\langle k|)=\sum_{j,k=1}^{d}\langle j|\rho|k\rangle T(|j\rangle\langle k|)\nonumber \\
 & = & \sum_{j,k=1}^{d}\langle j|\rho|k\rangle|k\rangle\langle j|=\sum_{j,k=1}^{d}\langle k|T(\rho)|j\rangle|k\rangle\langle j|.
\end{eqnarray}
Hence the familiar relation between the matrix elements of $\rho$
and of $T(\rho)$ is obtained, i.e., $\langle k|T(\rho)|j\rangle=\langle j|\rho|k\rangle$.

We remark that the definition in Eq. (\ref{eq:Tmap}) is base dependent.
So, for another basis $|\beta_{j}\rangle:=U|j\rangle,$ with $UU^{\dagger}=\mathbb{I}$
($\mathbb{I}$ is the identity operator in $\mathcal{H}$), the last
simple relation would be valid only for the ``rotated'' versions of
$T(\rho)$ and of $\rho$, i.e., $\langle\beta_{k}|(UT(\rho)U^{\dagger})|\beta_{j}\rangle=\langle\beta_{j}|(U\rho U^{\dagger})|\beta_{k}\rangle$.

Let's end this sub-section observing that once we have $\det(T(\rho-\lambda\mathbb{I}))=\det(\rho-\lambda\mathbb{I})$
\cite{Kuttler} and $T(\rho-\lambda\mathbb{I})=T(\rho)-\lambda\mathbb{I}$,
then the eigenvalues of $T(\rho)$ are the same as those of $\rho$.
An immediate consequence of this result is that their traces are also
equal, i.e., $\mathrm{Tr}(T(\rho))=\mathrm{Tr}(\rho)$. Thus, if $\rho$
is a density operator, i.e., it is positive semidefinite ($\rho\ge0$)
and has unit trace $(\mathrm{Tr}(\rho)=1$), then $T(\rho)$ is also
a valid density operator. This fact is key for the Peres' separability
criterion, which shall be recalled in Sec. \ref{Peres}.

\subsection{Partial transposition for bipartitions}

\label{sec:PT2}

In what follows we shall introduce the partial transposition (PT)
operation and obtain expressions which are useful for its numerical
implementation. Let's start regarding a bipartition of $\mathcal{H}$,
$\mathcal{H}_{a}\otimes\mathcal{H}_{b}$, with dimensions $d_{s}:=\dim\mathcal{H}_{s}$
for $s=a\mbox{, }b$. Any computational base state in $\mathcal{H}$
can be cast in terms of the local computational bases as follows:
$|j\rangle=|j_{a}\rangle\otimes|j_{b}\rangle=|j_{a}j_{b}\rangle=|(j_{a}-1)d_{b}+j_{b}\rangle,$
with $|j_{s}\rangle$ being the computational basis in $\mathcal{H}_{s}$.
Hereafter, we assume the the matrix elements of $\rho$ in the product-local
computational basis $|j_{a}j_{b}\rangle$ are known:
\begin{eqnarray}
\rho & = & \sum_{j_{a},k_{a}=1}^{d_{a}}\sum_{j_{b},k_{b}=1}^{d_{b}}\langle j_{a}j_{b}|\rho|k_{a}k_{b}\rangle|j_{a}\rangle\langle k_{a}|\otimes|j_{b}\rangle\langle k_{b}|.
\end{eqnarray}

With this, we are ready to introduce the, also linear, \emph{partial
transposition} operator, which, when taken over sub-system $a$, is
defined by $T_{a}\equiv T\otimes id,$ with $id$ being the identity
map, i.e., $id(X)=X$ for all linear operator $X$ on $\mathcal{H}_{s}$.
So,
\begin{eqnarray}
 &  & T_{a}(\rho)=T\otimes id(\rho)\nonumber \\
 & = & \sum_{j_{a},k_{a}=1}^{d_{a}}\sum_{j_{b},k_{b}=1}^{d_{b}}\langle j_{a}j_{b}|\rho|k_{a}k_{b}\rangle T(|j_{a}\rangle\langle k_{a}|)\otimes id(|j_{b}\rangle\langle k_{b}|)\nonumber \\
 & = & {\textstyle \sum_{j_{a},k_{a}=1}^{d_{a}}}{\textstyle \sum_{j_{b},k_{b}=1}^{d_{b}}}\langle j_{a}j_{b}|\rho|k_{a}k_{b}\rangle|k_{a}\rangle\langle j_{a}|\otimes|j_{b}\rangle\langle k_{b}|\nonumber \\
 & = & {\textstyle \sum_{j_{a},k_{a}=1}^{d_{a}}}{\textstyle \sum_{j_{b},k_{b}=1}^{d_{b}}}\langle j_{a}j_{b}|\rho|k_{a}k_{b}\rangle|k_{a}j_{b}\rangle\langle j_{a}k_{b}|.\label{eq:left}
\end{eqnarray}

In an analogous manner, when applied to sub-system $b$ the partial
transpose leads to
\begin{eqnarray}
 &  & T_{b}(\rho)=id\otimes T(\rho)\nonumber \\
 & = & \sum_{j_{a},k_{a}=1}^{d_{a}}\sum_{j_{b},k_{b}=1}^{d_{b}}\langle j_{a}j_{b}|\rho|k_{a}k_{b}\rangle|j_{a}\rangle\langle k_{a}|\otimes T(|j_{b}\rangle\langle k_{b}|)\nonumber \\
 & = & {\textstyle \sum_{j_{a},k_{a}=1}^{d_{a}}}{\textstyle \sum_{j_{b},k_{b}=1}^{d_{b}}}\langle j_{a}j_{b}|\rho|k_{a}k_{b}\rangle|j_{a}\rangle\langle k_{a}|\otimes|k_{b}\rangle\langle j_{b}|\nonumber \\
 & = & {\textstyle \sum_{j_{a},k_{a}=1}^{d_{a}}}{\textstyle \sum_{j_{b},k_{b}=1}^{d_{b}}}\langle j_{a}j_{b}|\rho|k_{a}k_{b}\rangle|j_{a}k_{b}\rangle\langle k_{a}j_{b}|.\label{eq:right}
\end{eqnarray}

In terms of matrix elements we get
\begin{eqnarray}
\langle k_{a}j_{b}|T_{a}(\rho)|j_{a}k_{b}\rangle & = & \langle j_{a}j_{b}|\rho|k_{a}k_{b}\rangle,\\
\langle j_{a}k_{b}|T_{b}(\rho)|k_{a}j_{b}\rangle & = & \langle j_{a}j_{b}|\rho|k_{a}k_{b}\rangle.
\end{eqnarray}
For numerical calculations, with the notation $\rho^{T_{s}}=T_{s}(\rho)$,
we just set
\begin{eqnarray}
\rho^{T_{a}}((k_{a}-1)d_{b}+j_{b},(j_{a}-1)d_{b}+k_{b}) & = & \rho(\alpha,\beta),\\
\rho^{T_{b}}((j_{a}-1)d_{b}+k_{b},(k_{a}-1)d_{b}+j_{b}) & = & \rho(\alpha,\beta),
\end{eqnarray}
with $\alpha=(j_{a}-1)d_{b}+j_{b}$ and $\beta=(k_{a}-1)d_{b}+k_{b}$
for all $j_{s},k_{s}=1,\cdots,d_{s}$. The PT for bipartite systems
is returned by the subroutines \texttt{partial\_transpose\_s($d_{a}$,
$d_{b}$, $\rho$, $T_{s}(\rho)$)}, with $s=a,b$.

\subsection{Partial transposition for multipartitions}

\label{sec:PT3}

Let's consider a density operator $\rho$ in the Hilbert space $\mathcal{H}_{a}\otimes\mathcal{H}_{b}\otimes\mathcal{H}_{c}$:
\begin{equation}
\rho=\sum\langle j_{a}j_{b}j_{c}|\rho|k_{a}k_{b}k_{c}\rangle|j_{a}j_{b}j_{c}\rangle\langle k_{a}k_{b}k_{c}|,
\end{equation}
with $|j_{s}\rangle$ and $|k_{s}\rangle$ being the computational
base for $\mathcal{H}_{s}$ ($s=a,b,c$) and the sum is made over
all $j_{s}$'s and $k_{s}$'s, which run from $1$ to $d_{s}$. Analogously
to the previous calculations, we apply the definition for the partial
transposition over the inner sub-system,
\begin{equation}
T_{b}(\rho)=id\otimes T\otimes id(\rho),\label{eq:inner}
\end{equation}
to see that
\begin{equation}
\langle j_{a}k_{b}j_{c}|\rho^{T_{b}}|k_{a}j_{b}k_{c}\rangle=\langle j_{a}j_{b}j_{c}|\rho|k_{a}k_{b}k_{c}\rangle.
\end{equation}
For numerical calculations, we use
\begin{equation}
|xyz\rangle=|(x-1)d_{b}d_{c}+(y-1)d_{c}+z\rangle
\end{equation}
to directly relate the matrix elements of $\rho^{T_{b}}$ and of $\rho$
in the global computational basis. The subroutine provided to compute
the inner partial transposition map is: \texttt{partial\_transpose\_3($d_{a}$,
$d_{b}$, $d_{c}$, $\rho$, $T_{b}(\rho)$)}.

Now, given any multipartite state space $\cdots\otimes\mathcal{H}_{s-1}\otimes\mathcal{H}_{s}\otimes\mathcal{H}_{s+1}\otimes\cdots\otimes\mathcal{H}_{s'-1}\otimes\mathcal{H}_{s'}\otimes\mathcal{H}_{s'+1}\otimes\cdots,$
we notice that the partial transposition over the parties $s$ and
$s'$ can be composed as follows:
\begin{equation}
T_{ss'}(\rho)\equiv T_{s}\circ T_{s'}(\rho).
\end{equation}
With this, the partial transposition over an arbitrary number of subsystems
(with arbitrary finite dimensions), can be computed through the composition
of the left (Eq. (\ref{eq:left})), right (Eq. (\ref{eq:right})),
and inner (Eq. (\ref{eq:inner})) partial transpositions described
above. We also provide a subroutine, \texttt{partial\_transpose($d$,
$\rho$, $T_{p}(\rho)$, nss, di, ssys)}, which returns the partial
transposition in the general case. Regarding the arguments therein,
\texttt{nss} is the number of sub-systems, $d$ is the total dimension,
\texttt{di} is a vector containing the dimensions of the subsystems,
and \texttt{ssys} is a vector with components equal to $0$ or $1$
for those subsystems over which the PT shall or shall not be applied,
respectively. The dimension of \texttt{di} and \texttt{ssys} is equal
to \texttt{nss}.

\subsection{Peres' criterion and entanglement negativity}

\label{Peres}

In 1996, A. Peres \cite{Peres_PT} made the insightful observation
that if a state is separable, i.e., if it can be cast as
\begin{equation}
\sigma={\textstyle \sum_{j}}p_{j}\sigma_{j}^{a}\otimes\sigma_{j}^{b}
\end{equation}
with $p_{j}$ being a probability distribution and $\sigma_{j}^{s}$
being valid density operators for the sub-system $s$, then its PT,
\begin{equation}
\tilde{\sigma}=T_{b}(\sigma)={\textstyle \sum_{j}}p_{j}\sigma_{j}^{a}\otimes T(\sigma_{j}^{b})={\textstyle \sum_{j}}p_{j}\sigma_{j}^{a}\otimes\tilde{\sigma}_{j}^{b},
\end{equation}
is also a valid (and separable) state, because $\tilde{\sigma}_{j}^{b}=T(\sigma_{j}^{b})$
are valid density operators (see Sec. \ref{transp}) and the convex
combination of positive semidefinite matrices is also a positive semidefinite
matrix \cite{Horn}. So, $\tilde{\sigma}$ is a positive semidefinite
matrix. Therefore, if the PT of a generic density matrix $\rho$ is
negative, then this state has to be entangled. This fact indicates
that the sum of the absolute values of the negative eigenvalues of
the PT of a state would be a possible entanglement quantifier. Actually,
the entanglement negativity \cite{Werner_En},
\begin{equation}
E_{n}(\rho)=2^{-1}(||T_{b}(\rho)||_{tr}-1),
\end{equation}
is an entanglement function \cite{Horodecki_RMP}. In the last equation
$||X||_{tr}:=\mathrm{Tr}\sqrt{X^{\dagger}X}$ is the trace norm. The
Fortran function \texttt{negativity($d$, $T_{p}(\rho)$)} returns
$E_{n}$ once provided the PT of $\rho$ and its dimension. In order
to obtain the logarithmic negativity \cite{Werner_En,Plenio_Eln},
\begin{equation}
E_{ln}(\rho):=\log_{2}(2E_{n}(\rho)+1),
\end{equation}
just change the name of
the function to \texttt{log\_negativity}.

It was shown later that the Peres' condition is necessary and sufficient
only for systems with dimension up to six \cite{Horodecki_PT}. For
larger dimensions, there may exist entangled states with positive
PT \cite{Horodecki_Ebound}. As a matter of fact, there is no known
analytically computable entanglement measure for general states \cite{Horodecki_RMP}.
In the next section we'll consider another entanglement function which
is not an entanglement quantifier, but which may be a useful analytical
tool in several circumstances.

\section{Analytical formula for the Hilbert-Schmidt entanglement}

\label{sec:HSE}

In this section we shall obtain an analytical expression for the Hilbert-Schmidt
entanglement (HSE) and for the associated closest separable state.
Our approach is motivated by Ref. \cite{Verstraete_E2}, but here
we do not use Lagrange multipliers. With this our calculations gain
in clarity and avoid possible drawbacks of that method \cite{Jing_LM}. 

As mentioned in Sec. \ref{intro}, computing entanglement quantifiers
(EQs) for general states is a very complex task. So, one of the motivations
for studying entanglement functions, such as the HSE, is that the
insights gained while doing that can shed some light on how we can
effectively tackle the complicated optimizations problems involved
in the calculation of EQs. On the other hand, the consideration of
the HSE, in addition to the entanglement negativity (EN), is appealing
because of the geometrical nature of the first. For instance, contrary
to the EN, when computing the HSE we can get as a byproduct the closest
separable state. And this kind of information can be useful, for example,
for studying the geometrical aspects of the quantum state space and
as an initial ansatz for the calculation of EQs induced by other,
more faithful, distinguishability measures. 

Let's recall that the Hilbert-Schmidt (HS) norm of a matrix $A$ is
defined and given by
\begin{equation}
||A||_{hs}:=\sqrt{\mathrm{Tr}(A^{\dagger}A)}=\sqrt{{\textstyle \sum_{j,k}}|\langle j|A|k\rangle|^{2}}.
\end{equation}
The HSE of a state $\rho$ is then defined, using the HS distance,
as
\begin{equation}
E_{hs}(\rho):=\min_{\sigma}||\rho-\sigma||_{hs},
\end{equation}
with the minimization running over all separable states.

Since the HS norm is invariant under unitary transformations, i.e.,
it is base independent, we can use the computational basis to verify
that the HS distance does not change under taking the PT of its arguments:
\begin{eqnarray}
 &  & ||T_{b}(\rho-\sigma)||_{hs}^{2}\nonumber \\
 & = & {\textstyle \sum_{j_{a},k_{a}=1}^{d_{a}}}{\textstyle \sum_{j_{b},k_{b}=1}^{d_{b}}}|\langle j_{a}j_{b}|(T_{b}(\rho)-T_{b}(\sigma))|k_{a}k_{b}\rangle|^{2}\nonumber \\
 & = & \sum_{j_{a},k_{a}=1}^{d_{a}}\sum_{j_{b},k_{b}=1}^{d_{b}}|\langle j_{a}j_{b}|T_{b}(\rho)|k_{a}k_{b}\rangle-\langle j_{a}j_{b}|T_{b}(\sigma)|k_{a}k_{b}\rangle|^{2}\nonumber \\
 & = & {\textstyle \sum_{j_{a},k_{a}=1}^{d_{a}}}{\textstyle \sum_{j_{b},k_{b}=1}^{d_{b}}}|\langle j_{a}k_{b}|\rho|k_{a}j_{b}\rangle-\langle j_{a}k_{b}|\sigma|k_{a}j_{b}\rangle|^{2}\nonumber \\
 & = & {\textstyle \sum_{j_{a},k_{a}=1}^{d_{a}}}{\textstyle \sum_{k_{b},j_{b}=1}^{d_{b}}}|\langle j_{a}j_{b}|\rho|k_{a}k_{b}\rangle-\langle j_{a}j_{b}|\sigma|k_{a}k_{b}\rangle|^{2}\nonumber \\
 & = & {\textstyle \sum_{j_{a},k_{a}=1}^{d_{a}}}{\textstyle \sum_{j_{b},k_{b}=1}^{d_{b}}}|\langle j_{a}j_{b}|(\rho-\sigma)|k_{a}k_{b}\rangle|^{2}\nonumber \\
 & = & ||\rho-\sigma||_{hs}^{2}.
\end{eqnarray}
Thus, using this equivalence, we can write 
\begin{equation}
E_{hs}(\rho)=\min_{\sigma}||T_{b}(\rho)-T_{b}(\sigma)||_{hs}=:\min_{\tilde{\sigma}}||\rho^{T_{b}}-\tilde{\sigma}||_{hs}.
\end{equation}

In the sequence we use once more the invariance under unitaries of
the HS norm to see that:
\begin{equation}
E_{hs}(\rho)=\min_{\tilde{\sigma}}||U(\rho^{T_{b}}-\tilde{\sigma})U^{\dagger}||_{hs}=:\min_{\zeta}||D-\zeta||_{hs},\label{eq:HSE1}
\end{equation}
where, considering that $T_{b}(\rho)$ is an Hermitian operator: 
\begin{eqnarray}
 &  & (T_{b}(\rho))^{\dagger}=\sum_{j_{a},k_{a}=1}^{d_{a}}\sum_{j_{b},k_{b}=1}^{d_{b}}\langle j_{a}j_{b}|\rho|k_{a}k_{b}\rangle^{*}(|j_{a}k_{b}\rangle\langle k_{a}j_{b}|)^{\dagger}\nonumber \\
 &  & ={\textstyle \sum_{j_{a},k_{a}=1}^{d_{a}}}{\textstyle \sum_{j_{b},k_{b}=1}^{d_{b}}}\langle k_{a}k_{b}|\rho^{\dagger}|j_{a}j_{b}\rangle|k_{a}j_{b}\rangle\langle j_{a}k_{b}|\\
 &  & ={\textstyle \sum_{j_{a},k_{a}=1}^{d_{a}}}{\textstyle \sum_{j_{b},k_{b}=1}^{d_{b}}}\langle j_{a}j_{b}|\rho|k_{a}k_{b}\rangle|j_{a}k_{b}\rangle\langle k_{a}j_{b}|=T_{b}(\rho),\nonumber 
\end{eqnarray}
we assumed that $U$ diagonalizes $\rho^{T_{b}}$, i.e.,
\begin{equation}
U\rho^{T_{b}}U^{\dagger}=D:={\textstyle \sum_{j=1}^{d}}D_{j}|D_{j}\rangle\langle D_{j}|,\label{eq:Ded}
\end{equation}
with $d=d_{a}d_{b}$. Besides we defined the, in principle general
and possibly entangled, density operator:
\begin{equation}
\zeta=U\tilde{\sigma}U^{\dagger}.\label{eq:zeta}
\end{equation}

We remark at this point that once we find the optimal $\zeta$, let's
call it $\zeta^{\star}$, then, as
\begin{equation}
T_{b}(\tilde{\sigma})=T_{b}(T_{b}(\sigma))=\sigma,
\end{equation}
we have found also the optimal separable state:
\begin{equation}
\sigma^{\star}=T_{b}(\tilde{\sigma}^{\star})=T_{b}(U^{\dagger}\zeta^{\star}U).\label{eq:css_star}
\end{equation}

Following with the calculation of the HSE, we use Eqs. (\ref{eq:HSE1}),
(\ref{eq:Ded}), and (\ref{eq:zeta}) to write 
\begin{eqnarray}
 &  & E_{hs}(\rho)=\min_{\zeta}\sqrt{\mathrm{Tr}(D-\zeta)^{2}}\\
 & = & \min_{\zeta}\sqrt{{\textstyle \sum_{j=1}^{d}}\langle D_{j}|(D-\zeta){\textstyle \sum_{k=1}^{d}}|D_{k}\rangle\langle D_{k}|(D-\zeta)|D_{j}\rangle}\nonumber \\
 & = & \min_{\zeta}\left({\textstyle \sum_{j,k}}(\langle D_{j}|D|D_{k}\rangle-\langle D_{j}|\zeta|D_{k}\rangle)\right.\nonumber \\
 &  & \hspace{1em}\hspace{1em}\hspace{1em}\hspace{1em}\hspace{1em}\hspace{1em}\hspace{1em}\hspace{1em}\cdot\left.(\langle D_{k}|D|D_{j}\rangle-\langle D_{k}|\zeta|D_{j}\rangle)\right)^{1/2}\nonumber \\
 & = & \min_{\zeta}\sqrt{\sum_{j,k}(D_{j}\delta_{jk}-\langle D_{j}|\zeta|D_{k}\rangle)(D_{j}\delta_{jk}-\langle D_{k}|\zeta|D_{j}\rangle)}\nonumber \\
 & = & \min_{\zeta}\sqrt{\sum_{j=k}(D_{j}-\langle D_{j}|\zeta|D_{j}\rangle)^{2}+\sum_{j\ne k}|\langle D_{j}|\zeta|D_{k}\rangle|^{2}}.\nonumber 
\end{eqnarray}

From this last expression, we see that $E_{hs}$ is minimized if $\zeta$
has no coherences in the eigenbasis of $D$, i.e., if
\begin{equation}
\zeta={\textstyle \sum_{j=1}^{d}}\zeta_{j}|D_{j}\rangle\langle D_{j}|.\label{eq:zeta1}
\end{equation}
So,
\begin{equation}
E_{hs}(\rho)=\min_{\{\zeta_{j}\}_{j=1}^{d}}\sqrt{{\textstyle \sum_{j=1}^{d}}(D_{j}-\zeta_{j})^{2}}.\label{eq:HSE2}
\end{equation}

In what follows, it will be useful noticing that, as $\zeta$ is a
density operator, we have to have $\zeta_{j}\ge0$ and $\mathrm{Tr}(\zeta)=\sum_{j=1}^{d}\zeta_{j}=1$.
In addition to that, it will be important for our calculations seeing
that $T_{b}(\rho)$ has unit trace:
\begin{eqnarray}
 &  & \mathrm{Tr}(T_{b}(\rho))=\sum_{j_{a},k_{a}=1}^{d_{a}}\sum_{j_{b},k_{b}=1}^{d_{b}}\langle j_{a}j_{b}|\rho|k_{a}k_{b}\rangle\mathrm{Tr}(|j_{a}k_{b}\rangle\langle k_{a}j_{b}|)\nonumber \\
 &  & ={\textstyle \sum_{j_{a},k_{a}=1}^{d_{a}}}{\textstyle \sum_{j_{b},k_{b}=1}^{d_{b}}}\langle j_{a}j_{b}|\rho|k_{a}k_{b}\rangle\delta_{j_{a}k_{a}}\delta_{k_{b}j_{b}}\nonumber \\
 &  & ={\textstyle \sum_{j_{a}=1}^{d_{a}}}{\textstyle \sum_{j_{b}=1}^{d_{b}}}\langle j_{a}j_{b}|\rho|j_{a}j_{b}\rangle=\mathrm{Tr}(\rho)=1.
\end{eqnarray}

Now, let $D_{j}^{+}$, $D_{j}^{-}$, and $D_{j}^{0}$ denote the (real)
positive, negative, and null eigenvalues of $D$ (and of $T_{b}(\rho))$.
The dimension of the corresponding eigenspaces are denoted, respectively,
by $d_{+}$, $d_{-}$, and $d_{0}$; thus $d_{+}+d_{-}+d_{0}=d$.
Hence, the unit trace of $T_{b}(\rho)$ leads to:
\begin{eqnarray}
 &  & \mathrm{Tr}(\rho^{T_{b}})=1=\mathrm{Tr}(U\rho^{T_{b}}U^{\dagger})=\mathrm{Tr}(D)={\textstyle \sum_{j=1}^{d}}D_{j}\\
 &  & =\sum_{j=1}^{d_{+}}D_{j}^{+}+\sum_{j=1}^{d_{-}}D_{j}^{-}+\sum_{j=1}^{d_{0}}0=\sum_{j=1}^{d_{+}}D_{j}^{+}-\sum_{j=1}^{d_{-}}|D_{j}^{-}|,\nonumber 
\end{eqnarray}
which is obtained only if $\sum_{j=1}^{d_{+}}D_{j}^{+}\ge1$. Thus,
the HSE in Eq. (\ref{eq:HSE2}) can be written as
\begin{eqnarray}
E_{hs}(\rho) & = & \min_{\{\zeta_{j}\}_{j=1}^{d}}\left({\textstyle \sum_{j=1}^{d_{+}}}(D_{j}^{+}-\zeta_{j}^{+})^{2}\right.\hspace{1em}\hspace{1em}\label{eq:HSE3}\\
 &  & \hspace{1em}\left.+{\textstyle \sum_{j=1}^{d_{-}}}(D_{j}^{-}-\zeta_{j}^{-})^{2}+{\textstyle \sum_{j=1}^{d_{0}}}(0-\zeta_{j}^{0})^{2}\right)^{1/2},\nonumber 
\end{eqnarray}
where $\zeta_{j}^{+},\zeta_{j}^{-}\mbox{, and }\zeta_{j}^{0}$ are
the eigenvalues of $\zeta$ with eigenvectors in the positive, negative,
and null eigenspaces of $D$, respectively. Of course, we shall minimize
$E_{hs}$ if we set $\zeta_{j}^{0}:=0$ for $j=1,\cdots,d_{0}$. We
also minimize $E_{hs}$ if we set $\zeta_{j}^{-}:=0$ for $j=1,\cdots,d_{-}$
(because any $\zeta_{j}^{-}>0$ would make $D_{j}^{-}-\zeta_{j}^{-}$
more negative and hence lead to a greater value of $(D_{j}^{-}-\zeta_{j}^{-})^{2}$).
Next, let the positive eigenvalues $D_{j}^{+}$ be arranged in decreasing
order and let $d_{+}'$ be defined such that
\begin{equation}
1-\xi:={\textstyle \sum_{j=1}^{d_{+}'-1}}D_{j}^{+}\le1\mbox{ and }{\textstyle \sum_{j=1}^{d_{+}'}}D_{j}^{+}>1.
\end{equation}
Then, considering that S. Rana showed in Ref. \cite{Rana} that for
two-qudit states the eigenvalues of the PT of $\rho$ lie in interval
$[-1/2,1]$, we shall minimize $E_{hs}$ if we set
\begin{eqnarray}
 &  & \zeta_{j}^{+}=D_{j}^{+}\mbox{ for }j=1,\cdots,d_{+}'-1,\nonumber \\
 &  & \zeta_{d_{+}'}^{+}=\xi,\\
 &  & \zeta_{j}^{+}=0\mbox{ for }j=d_{+}'+1,\cdots,d_{+}.\nonumber 
\end{eqnarray}
With these choices for $\zeta_{j}$, we'll have a valid density operator
$\zeta$. Thus, substituting these values of $\zeta_{j}$ in Eq. (\ref{eq:HSE3}),
the minimum value for the \emph{Hilbert-Schmidt entanglement} of an
arbitrary bipartite density matrix shall be given by:
\begin{eqnarray}
E_{hs}^{2}(\rho) & = & (D_{d_{+}'}^{+}-\xi)^{2}+{\textstyle \sum_{j=d_{+}'+1}^{d_{+}}}(D_{j}^{+})^{2}+{\textstyle \sum_{j=1}^{d_{-}}}(D_{j}^{-})^{2}\nonumber \\
 & = & \left(\sum_{j=1}^{d_{+}'}D_{j}^{+}-1\right)^{2}+\sum_{j=d_{+}'+1}^{d_{+}}(D_{j}^{+})^{2}+\sum_{j=1}^{d_{-}}(D_{j}^{-})^{2}\nonumber \\
 & = & \left({\textstyle \sum_{j=1}^{d_{-}}}|D_{j}^{-}|-{\textstyle \sum_{j=d_{+}'+1}^{d_{+}}}D_{j}^{+}\right)^{2}\nonumber \\
 &  & +{\textstyle \sum_{j=d_{+}'+1}^{d_{+}}}(D_{j}^{+})^{2}+{\textstyle \sum_{j=1}^{d_{-}}}(D_{j}^{-})^{2}.\label{eq:HSE}
\end{eqnarray}

We observe that $E_{hs}$ is written above in terms of (all) the negative
eigenvalues of $T_{b}(\rho)$ and in terms of its $d_{+}-d_{+}'$
smaller positive eigenvalues. If the state under analysis is separable,
then, in addition to the eigenvalues of $T_{b}(\rho)$ being positive
(i.e., $d_{-}=0$), we have $d_{+}=d_{+}'-1$ and therefore $E_{hs}(\sigma)=0$,
as expected. Besides, $E_{hs}(\rho)>0$ whenever $E_{n}(\rho)>0$.

To obtain the \emph{closest separable state} (CSS), we start using
the optimal $\zeta_{j}$'s to write Eq. (\ref{eq:zeta1}) as follows:
\begin{eqnarray}
\zeta^{\star} & = & {\textstyle \sum_{j=1}^{d}}\zeta_{j}^{\star}|D_{j}\rangle\langle D_{j}|\\
 & = & {\textstyle \sum_{j=1}^{d_{+}'-1}}D_{j}^{+}|D_{j}^{+}\rangle\langle D_{j}^{+}|+\xi|D_{d_{+}'}^{+}\rangle\langle D_{d_{+}'}^{+}|\nonumber \\
 &  & +{\textstyle \sum_{j=d_{+}'+1}^{d_{+}}}0|D_{j}^{+}\rangle\langle D_{j}^{+}|+{\textstyle \sum_{j=1}^{d_{-}}}0|D_{j}^{-}\rangle\langle D_{j}^{-}|\nonumber \\
 &  & +{\textstyle \sum_{j=1}^{d_{0}}}0|D_{j}^{0}\rangle\langle D_{j}^{0}|\nonumber \\
 & = & {\textstyle \sum_{j=1}^{d_{+}'-1}}D_{j}^{+}|D_{j}^{+}\rangle\langle D_{j}^{+}|+\xi|D_{d_{+}'}^{+}\rangle\langle D_{d_{+}'}^{+}|.\nonumber 
\end{eqnarray}
Thus, using Eq. (\ref{eq:css_star}) and noticing from Eq. (\ref{eq:Ded})
that if $T_{b}(\rho):=\sum_{j=1}^{d}D_{j}|R_{j}\rangle\langle R_{j}|$
then $|R_{j}\rangle=U^{\dagger}|D_{j}\rangle$, we get
\begin{equation}
\sigma^{\star}=T_{b}({\textstyle \sum_{j=1}^{d_{+}'-1}}D_{j}^{+}|R_{j}^{+}\rangle\langle R_{j}^{+}|+\xi|R_{d_{+}'}^{+}\rangle\langle R_{d_{+}'}^{+}|)=:T_{b}(\Xi).\label{eq:css}
\end{equation}
So, as $|R_{j}^{+}\rangle$ is the eigenvector of $T_{b}(\rho)$ corresponding
to its $j$-th positive eigenvalue, we have written $\sigma^{\star}$
in terms of quantities directly related to the PT of $\rho$. Actually,
the closest separable state from $\rho$ is seem to be the PT of the
mixture of the $d_{+}'$ eigenvectors of $T_{b}(\rho)$ corresponding
to its $d_{+}'$ greater eigenvalues; with the weights given by the
eigenvalues themselves or by $\xi$. 

\begin{figure}[b]
\begin{centering}
\includegraphics[scale=0.4]{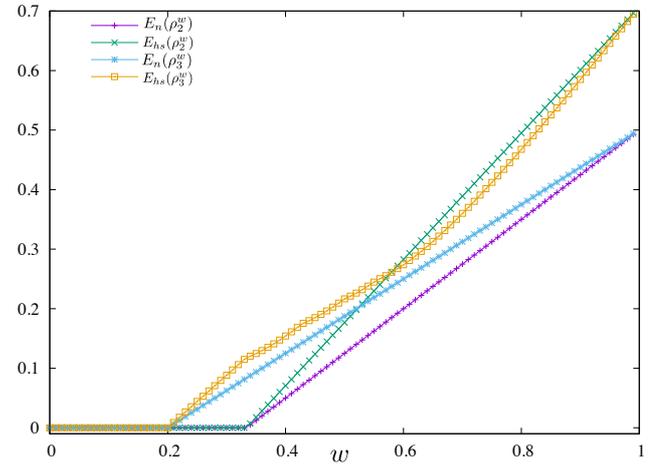}
\par\end{centering}

\caption{(color online) Entanglement negativity and Hilbert-Schmidt entanglement
for the two- and three-qubit Werner states of Eq. (\ref{eq:werner}).
These states are fully separable for $w$ less than $1/3$ and $1/5$,
respectively \cite{Rubin}. In the two cases we apply the partial
transposition to one of the qubits, the other two are regarded as
a ququart. For $\rho_{2}^{w}$ the positive eigenvalues of its partial
transpose are not used and the increasing rate of $E_{hs}$ with $w$
is constant. However, for $\rho_{3}^{w}$, because of the changes
of $d_{+}'$ with $w$, the number of positive eigenvalues involved
in the calculation of $E_{hs}$ also changes, and this leads to the
behavior shown in the plot.}

\label{fig:werner}
\end{figure}

The HSE, Eq. (\ref{eq:HSE}), and the matrix $\Xi$ in Eq. (\ref{eq:css}),
whose PT gives the CSS, are returned by the subroutine \texttt{entanglement\_hs($d$,
$T_{b}(\rho)$, $E_{hs}$, css)}, with \texttt{css} being a \texttt{character(1)}
variable. If \texttt{css = `y'} then, on exit, $\Xi$ is returned
in $T_{b}(\rho)$. If \texttt{css = `n'} and/or $E_{hs}(\rho)=0$
then $\Xi$ is not computed and $T_{b}(\rho)$ is not modified. As
an example, in Fig. \ref{fig:werner} we show $E_{n}$ and $E_{hs}$
calculated for the two- and three-qubit Werner states:
\begin{equation}
\rho_{n}^{w}=w|\Phi_{n}\rangle\langle\Phi_{n}|+(1-w)2^{-n}\mathbb{I}_{2^{n}},\label{eq:werner}
\end{equation}
where $w\in[0,1]$, $|\Phi_{2}\rangle=2^{-1/2}(|00\rangle+|11\rangle)$,
$|\Phi_{3}\rangle=2^{-1/2}(|000\rangle+|111\rangle)$, and $\mathbb{I}_{2^{n}}$
is the $2^{n}\mathrm{x}2^{n}$ identity matrix.

\section{Concluding remarks}

\label{conc}

In this article, we presented a thorough description of the partial
transposition (PT) map and of related entanglement functions. We produced
and described free Fortran code to compute all of these functions.
Besides, considering two-qudit systems, we obtained an analytical
expression for the Hilbert-Schmidt entanglement (HSE) and for the
associated nearest separable state. In our derivation, we used basically
the properties of the PT of a state. So, in addition to its simplicity
and clarity, our approach may be more suitable when compared to the
application of Lagrange multipliers \cite{Jing_LM}.

It is worthwhile remarking that the HS distance (HSD) is not generally
contractive under quantum operations \cite{Petz_contractivity,Wang_contractivity}.
This fact has motivated critiques regarding its use for quantum correlations
quantification \cite{Ozawa_HS,Piani_contractivity}. Although the
HSD can still be a formidable tool for several kinds of inquires \cite{Vedral_RSP1,Bertlmann0,Bertlmann1,Brukner_HSD,Cohen_HSD,Dodonov_HSD,Zyczkowski_HSV,Popescu_E_gibbs},
it would be interesting verifying if the procedure presented here
to compute the HSE can be extended to other distance measures possessing
more of the wanted ``good'' properties. In this direction, it is interesting
observing that the $l_{1}$-norm, which when computed using the basis
$\mathcal{B}=\{|b_{j}\rangle\}$ is given by $||A||_{l_{1}}^{\mathcal{B}}=\sum_{j,k}|\langle b_{j}|A|b_{k}\rangle|$,
was shown to lead to a faithful quantum coherence quantifier (in contrast
to $||A||_{hs}$) \cite{Plenio_Coh}. In fact, if $\mathcal{B}$ is
the computational basis: $\mathcal{C}=\{|j_{a}j_{b}\rangle\}$, we
can show, in an analogous manner to the verification in Sec. \ref{sec:HSE},
that
\begin{equation}
||\rho-\sigma||_{l_{1}}^{\mathcal{C}}=||T_{b}(\rho-\sigma)||_{l_{1}}^{\mathcal{C}}.
\end{equation}
However, the lack of unitary invariance of the $l_{1}$-norm \cite{Adesso_L1}
seems to complicate its application in this scenario; so we leave
the verification of this possibility as an open problem. As an alternative,
it would be interesting considering also the $r_{1}$-norm, introduced
in Ref. \cite{r1_norm}, and its quantum extension for application
in this context. It remains though to be investigated if the induced
distance measure retains the properties of unitary-invariance, computability,
PT-invariance, and contractivity under quantum operations.

\begin{acknowledgements}
This work was supported by the Brazilian funding agencies: Conselho Nacional de Desenvolvimento Cient\'ifico e Tecnol\'ogico (CNPq), processes 441875/2014-9 and 303496/2014-2, Instituto Nacional de Ci\^encia e Tecnologia de Informa\c{c}\~ao Qu\^antica (INCT-IQ), process 2008/57856-6, and Coordena\c{c}\~ao de Desenvolvimento de Pessoal de N\'{i}vel Superior (CAPES), process 6531/2014-08. I thank the hospitality of the Physics Institute and Laser Spectroscopy Group at the Universidad de la Rep\'{u}blica, Uruguay. I also thank Adriana Auyuanet for bringing Ref. \cite{Verstraete_E2} to my attention.
\end{acknowledgements}

\end{document}